\documentclass[twocolumn,aps,superscriptaddress,showpacs]{revtex4}
\usepackage{amsmath,bm}
\usepackage{graphicx}
\begin{document}

\draft
%\preprint{}
\title{Comparison of elastic electron or positron scattering
from proton-rich nuclei}

\thanks{Supported partially by the Shanghai Development Foundation from Science
and Technology under Grant Numbers 05XD14021 and 06QA14062, the
National Natural Science Foundation of China under Grant No
10328259, 10135030, and 10535010, and the Major State Basic
Research Development Program under Contract No G200077404.}

\author{E. J. MA}
\affiliation{\footnotesize Shanghai Institute of Applied Physics,
Chinese Academy of Sciences, P. O. Box 800-204, Shanghai 201800}
\affiliation{\footnotesize Changshu Institute of Technology,
Jiangsu 215500} \affiliation{\footnotesize Graduate school of
Chinese Academy of Sciences, Beijing 100039}
\author{Y. G. MA}
\thanks{Corresponding author. Email: ygma@sinap.ac.cn}
\author{J. G. CHEN}
\author{X. Z. CAI}
\author{D. Q. FANG}
\author{W. GUO}
\author{G. H. LIU}
\author{C. W. MA}
\author{W. Q. SHEN}
\author{Y. SHI}
\author{Q. M. SU}
\author{W. D. TIAN}
\author{H. W. WANG}
\author{K. WANG}
\author{T. Z. YAN}
\affiliation{\footnotesize Shanghai Institute of Applied Physics,
Chinese Academy of Sciences, P. O. Box 800-204, Shanghai 201800}

%\email{}

%\address{}
\date{\today}
\nopagebreak

\begin{abstract}
We investigate the cross sections of the elastic electron or
positron scattering from $^{208}$Pb, $^{12}$C, $^{12,16}$O and
$^{28, 32}$S by the relativistic partial-wave expansion method
using the static charge density distribution from the
self-consistent relativistic mean field model and also calculate
the charge form factor for $^{12, 16}$O and $^{28, 32}$S. The
numerical results are compared with the available data.
Calculations indicate that the extended charge density
distributions of $^{12}$O and $^{28}$S have observable effects on
the cross sections of the electron or positron scattering as well
as the charge form factors.

\end{abstract}


%\keywords{}

\maketitle

The use of radioactive nuclear beams (RNB)  led to the discovery
of neutron halo nuclei \cite{Tanihata85}. Besides neutron halo,
another topic on halo nuclei is to search for proton halo from
proton-rich nuclei. Theoretically, many calculations have been
made to predict that there may be proton halos in $^{26,27,28}$P
\cite{ZRen96}, $^{23}$Al, $^{13}$N \cite{Zhang03} etc. Although
some measured data from experiments indicate some evidences of the
existence of proton halos in these nuclei \cite{Navin98}, further
experiments are needed to confirm the existence of the proton
halos. It is well known that the electron scattering off nuclei
provides the most direct information about charge distribution.
Technical proposals for an electron-heavy-ion collider has been
incorporated in the GSI/Germany physics program \cite{Haik05} as
well as the RIKEN/Japan facility \cite{Suda04}. In both cases the
main purpose is to study the structure of nuclei far from the
stability line. Since the electron-nucleus scattering is a better
way for the precise investigation of the extended charge
distribution and the experimental data for the electron scattering
off unstable nuclei will be available soon, it is of interesting
to make an exploratory investigation of elastic electron
scattering from proton-rich nuclei.

There are some methods which can calculate the differential cross
sections for the elastic high-energy electron scattering off
nuclei. The conventional methods are the distorted wave Born
approximation \cite{Antonov05}, the relativistic eikonal
approximation \cite{ZWang04} and the relativistic partial-wave
expansion method \cite{Yennie54}. The partial-wave expansion
method is generally believed to be an \lq\lq exact" method if
enough partial waves are included in the phase-shift calculations.
This method can deal with both electron and positron elastic
scattering off nuclei, so we utilize it to obtain the differential
cross sections and form factors for electron or positron
scattering off the exotic nuclei.

Some theoretical models can give reliable charge density
distribution for stable nuclei. The typical models include the
Skyrme-Hartree-Fork model \cite{Richter03}, the ab initio no-core
shell model \cite{Hasan04}, the large-scale shell model method
\cite{Karat97} and the self-consistent relativistic mean field
(RMF) model. A series of calculations \cite{Gambhir90, Horowitz81,
ZMa94, ZRen99} show that the RMF model can reproduce with good
precision the binding energies, the separation energies and the
radii of nuclear charge density distribution. Therefore we use the
RMF model to calculate the nuclear ground-state charge
distribution.

The core of the relativistic partial expansion method is to solve
a Dirac equation for the electron or positron in which the effect
of the nucleus is contained entirely in the static Coulomb
potential. We assume that the charge distribution of the target
nucleus is spherically symmetric. The electrostatic interaction
energy between the projectile and the target nucleus is
\begin{equation}\label{}
    V(r) = Z_{0}e\varphi(r),
\end{equation}
where $Z_{0}e$ is the charge of the projectile ($Z_{0} =-1 $ for
electron and $+1$ for positron) and $\varphi(r)$  is the
electrostatic potential of the target nucleus. Let $\rho_{ch}(r)$
denotes the charge density of the target nucleus, we have
\begin{equation}\label{Eq_NSP}
    \varphi(r)=e\left(\frac{1}{r}\int_{0}^{r} \rho_{ch}(r')4\pi r'^{2}dr'+
    \int_{r}^{\infty}\rho_{ch}(r')4\pi r'dr'\right).
\end{equation}
The scattering of relativistic electrons or positrons by a central
field $V(r)$ is completely described by the direct scattering
amplitude $f(\theta)$ and the spin-flip scattering amplitude
$g(\theta)$ \cite{Walker71}. The elastic differential cross
section in the relativistic partial-wave expansion method can be
expressed as
\begin{equation}\label{Eq_DCS}
    \frac{d\sigma}{d\Omega}=|f(\theta)|^{2}+|g(\theta)|^{2}.
\end{equation}
After the differential cross sections are obtained, we can
calculate the charge form factors $F(q)$ by dividing the
differential cross sections with the Mott cross section
$(d\sigma/d\Omega)_{Mott}$ \cite{Hofstadter56}
\begin{equation}\label{Eq_SCFF}
    |F(q)|^2=\frac{d\sigma/d\Omega}
    {(d\sigma/d\Omega)_{Mott}}.
\end{equation}
Starting from Eq.(\ref{Eq_NSP}), one can use Eqs.(\ref{Eq_DCS})
and (\ref{Eq_SCFF}) to predict the differential cross section and
the form factor for a given nuclear charge density distribution
$\rho_{ch}(r)$.

To investigate the validity of the relativistic partial-wave
method and the RMF model, we calculated positron or electron
elastic scattering cross sections and their ratios for $^{208}$Pb
and $^{12}$C for which the experimental data are available.
\begin{figure}[h]
  \includegraphics[width=9.0cm,angle=0.]{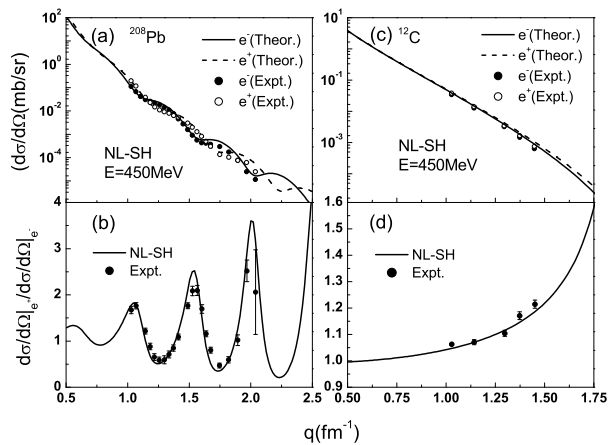}\\
  \caption{Positron and electron elastic scattering cross sections
  and their ratios for $^{208}$Pb (left column) and $^{12}$C (right column).}\label{Pb-C}
\end{figure}
In Fig.\ref{Pb-C}, we show comparisons of electron and positron
elastic scattering cross sections with the experimental ones
\cite{Breton91} for $^{208}$Pb and $^{12}$C using a static charge
density calculated by the RMF model with NL-SH parameters. The
electron and positron are all leptons, but their electromagnetic
interactions with nuclei are repulsive and attractive,
respectively. So the Coulomb distortion of the electron or
positron wave function by the scattering potential is different.
The part of the Coulomb effect associated with the acceleration of
the lepton by the Coulomb potential of the nucleus corresponds
\cite{Yennie65} to the effective momentum transferred between the
lepton and the nucleus depending on the sign of the lepton's
charge; this shifts the apparent location of the diffraction
minima. The effects of the Coulomb distortion and the different
effective momentum transfer are clearly visible in
Fig.\ref{Pb-C}(a); the locations of the diffraction minima are
shifted between electrons and positrons due to the change in the
effective momentum transfer, and the magnitudes of the cross
sections at the same effective momentum transfer differ due to
differences in the Coulomb distortion. Clearly, an
electron-positron comparison is highly sensitive to the Coulomb
distortion effects, and the calculation is in general agreement
with the measured data.
 Fig.\ref{Pb-C} also displays the theoretical cross sections and their ratios for $^{12}$C
 compared to the experimental data. The good agreement between experimental results and
 theoretical calculations for the $^{12}$C data is also reached.

For each incident energy and momentum transfer we determined the
ratio $R=(d\sigma/\Omega)|_{e^{+}}/(d\sigma/\Omega)|_{e^{-}}$ for
both the measured data and the theoretical calculation.
Fig.\ref{Pb-C}(b) (d) displays theoretical and experimental values
of $R$ for $^{208}$Pb and $^{12}C$. The theoretical ratios are
generally in good agreement with the measured ones\cite{Breton91}.

 From Fig.\ref{Pb-C}, it indicates that the combination of the
 relativistic partial-wave method with the RMF model can give reasonable  cross sections
 of high-energy electron or positron scattering off nuclei and the RMF model
 can produce the ground state charge density distribution of nuclei.
\begin{figure}[htbp]
  \includegraphics[width=9.0cm,angle=0.]{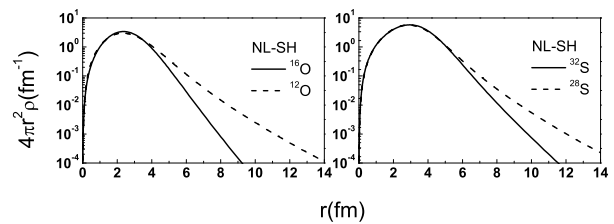}\\
  \caption{The charge density distribution of $^{12, 16}$O and
  $^{28, 32}$S calculated with NL-SH force parameters in RMF model.}\label{DEN}
\end{figure}

The charge density distributions of $^{12,16}$O and $^{28,32}$S
are displayed in Fig.\ref{DEN}. It is seen from Fig.\ref{DEN} that
the charge density distributions of $^{28,32}$S with NL-SH
parameters are different although the two nuclei have the same
proton number. The weak binding of the last two protons in
$^{28}$S leads to the extended charge density distribution in it.
This agrees with the theoretical calculations \cite{ZRen96},
\cite{ZRen99} and with the experimental results
 \cite{Navin98} of the neighboring nuclei $^{26, 27, 28}$P.
 The similar conclusion holds true for $^{12,16}$O.

\begin{figure}[hbtp]
  \includegraphics[width=9.0cm,angle=0.]{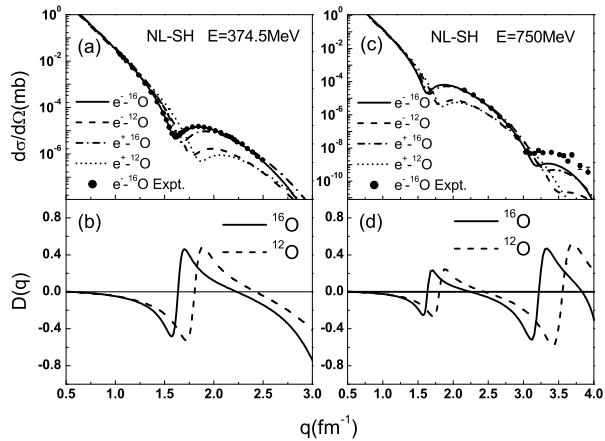}\\
  \caption{Positron and electron elastic scattering cross sections
  and their cross section differences for $^{12, 16}$O.}\label{O3775}
\end{figure}

\begin{figure}[hbtp]
  \includegraphics[width=9.0cm,angle=0.]{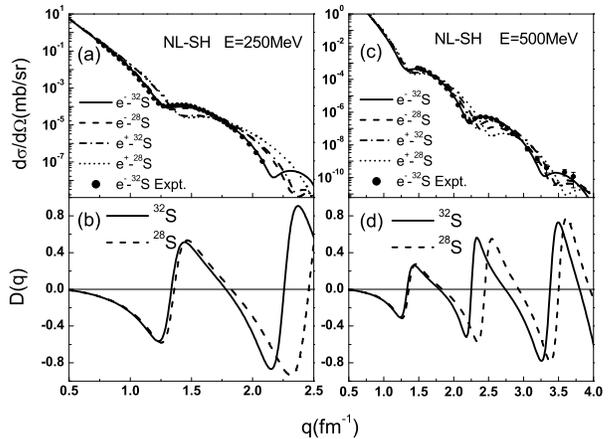}\\
  \caption{Positron and electron elastic scattering cross
  sections and their cross section differences  for  $^{28, 32}$S.
  }\label{S2550}
\end{figure}

In order to investigate the effect of the long tail of the charge
density distribution of proton-rich nuclei on the process of
elastic electron or positron scattering, the elastic scattering
cross sections for the proton-rich nuclei $^{28}$S and $^{12}$O
and respective stable isotopes $^{32}$S and $^{16}$O were
calculated with the partial-wave method. Fig.\ref{O3775} displays
the positron and electron elastic scattering cross sections and
their cross section differences for $^{12}$O and $^{16}$O at
different incident energies. The cross sections of the positron
and electron elastic scattering off $^{28, 32}$S for different
incident energies are plotted in Fig.\ref{S2550}. The difference
of the cross sections $D(q)$ in those figures is defined as
\begin{equation}\label{}
    D(q)=\frac{(d\sigma/d\Omega)|_{e^-}-(d\sigma/d\Omega)|_{e^+}}
    {(d\sigma/d\Omega)|_{e^-}+(d\sigma/d\Omega)|_{e^+}}
\end{equation}
for electron- and positron-scattering from the identical nucleus.
From those two figures, we can deduce the following conclusions.

Firstly, it is apparent that the calculated cross sections of
electron scattering both for $^{16}$O and for $^{32}$S almost
coincide with the experimental data \cite{Sick70,GCLi74} in the
range of low and moderate momentum transfer ($q\leq$3fm$^{-1}$).
Theoretical results have very good agreements with the
experimental data in this range of momentum transfer. At
high-momentum transfers, a deviation occurs between the
theoretical cross sections and the experimental data. Since the
cross section in this range of momentum transfer is mainly
sensitive to the details of the inner part of the charge density
distribution \cite{Sick70}, its occurrence indicates that the
theoretical charge density distribution has a departure from the
experimental one around the center of the nucleus. This means that
the RMF model can reproduce the charge density distribution for
$^{16}$O and $^{32}$S very well except near the center of the
nucleus. The combination of the relativistic partial-wave method
with the RMF model can predict the cross sections of the elastic
electron-nucleus scattering from the stable isotopes of O and S.

Secondly, the cross sections of electron scattering for the
unstable proton-rich nuclei $^{12}$O and $^{28}$S display the
outward shifts of the positions of the diffraction minima as
compared to the stable isotopes $^{16}$O and $^{32}$S. The
differences of the cross sections both for $^{12}$O and $^{16}$O
at the first minima of momentum transfer and for $^{32}$S and
$^{28}$S at the second minima of momentum transfer are large
enough to be observable in the experiments. On the one side, it
can be deduced from the fitting to the experimental data of
$^{12}$C \cite{Sick70} and $^{32}$S \cite{GCLi74} that the cross
sections in the range of moderate-momentum transfers,
approximately $1fm^{-1} \leq q\leq 3fm^{-1}$, are sensitive to the
tail of the nuclear charge distribution. On the other side, the
weak binding of the last two protons in $^{12}$O and $^{28}$S
leads to the extended charge density distribution  and the cross
section of the elastic electron-nucleus scattering is related to
the nuclear charge distribution. So the information of the long
tail of the charge density distribution of $^{12}$O and $^{28}$S
can be deduced by measurement of the cross sections of electron
scattering from unstable proton-rich nuclei $^{12}$O and $^{28}$S
at moderate-momentum transfers. Since the difference of the charge
density distributions both for $^{12}$O and $^{16}$O and for
$^{28}$S and $^{32}$S is mainly caused by the difference of the
charge density distributions of the last two protons in $^{12,
16}$O or $^{28, 30}$S, the information of the density distribution
of the last two protons could be extracted by comparing the cross
sections of $^{12, 16}$O or of $^{28, 30}$S.

Thirdly, the cross sections of positron scattering from the stable
nuclei $^{16}$O and $^{32}$S have the similar trends as compared
with the ones of electron scattering from the same nucleus. Like
the electron scattering from the unstable nuclei $^{12}$O and
$^{28}$S, the positions of the diffraction minima of the cross
sections of positron scattering from the unstable nuclei $^{12}$O
and $^{28}$S shift outward in comparison to the positron
scattering from the stable nuclei $^{16}$O and $^{32}$S. The
difference between the cross section of positron scattering off
the stable nucleus and the corresponding unstable isotope is
apparent to be distinguishable in the positron-nucleus scattering
experiment. So one can presume that the conclusions of analysis of
electron-nucleus scattering still hold true for the
positron-nucleus scattering, and the measurement and comparison of
the cross sections of the positron scattering from unstable
nucleus and the corresponding stable nucleus can provide the
information of the ground-state charge density distribution of
unstable  proton-rich nuclei.
\begin{figure}[hbtp]
  \includegraphics[width=9.0cm,angle=0.]{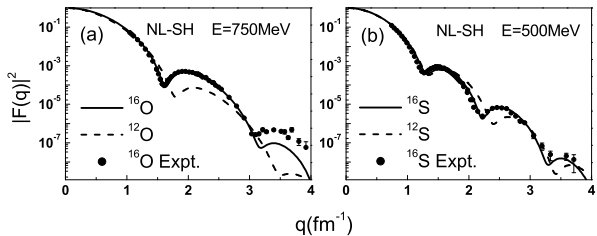}\\
  \caption{The squared charge form factor for $^{12, 16}$O and $^{28, 32}$S.
  }\label{OS-CFF}
\end{figure}

Fourthly, from analysis of the variation of the differences of the
cross sections $D(q)$ with momentum transfer, one can deduce that
the locations of diffraction minima of the cross section for the
positron-nucleus scattering are shifted as compared with the
electron-nucleus scattering. This shift is due to the different
Coulomb effect of the target nucleus on electron and positron. The
differences of the cross sections $D(q)$ oscillate irregularly and
the amplitudes of $D(q)$ increase with momentum transfer. In the
range of low momentum transfers, the solid curve (for $^{16}$O and
$^{32}$S) coincides with the dashed curve (for $^{12}$O and
$^{28}$S). This indicates that the differences of the cross
sections between stable nucleus and corresponding unstable isotope
can be observed in electron- or positron- scattering experiments
only when the momentum transfers are larger than a certain value.

For the purpose of comparing the experimental data of
electron-nucleus scattering at different incident energies, we
calculated the charge form factor of $^{12, 16}$O and $^{28,
32}$S. Fig.\ref{OS-CFF} displays the squared form factors of
$^{12, 16}$O and of $^{28, 32}$S. The experimental cross sections
for $^{16}$O and $^{32}$S are taken from Refs
\cite{Sick70,GCLi74}. The experimental cross sections at two
different incident energies have been transformed into the squared
form factors. It can be seen from Fig.\ref{OS-CFF} that the
location of the first minimum of the form factor of $^{12}$O and
the position of the second minimum of the form factor of $^{28}$S
shift outward markedly as compared with $^{16}$O and $^{32}$S. In
addition, the amplitude of the form factor has a significant
deviation, especially in the neighborhood of the two minima. The
shifts of the two minima and the amplitude deviation of the form
factor are large enough to be observable. Since the form factor is
connected to the charge density distribution by a Fourier
transformation, the elastic electron scattering form factor of a
nucleus is directly related to its charge density distribution
which is mainly determined by the density distribution of protons
in a nucleus. The weak-binding of the last two protons in $^{12}$O
and $^{32}$S leads to the extended charge density distribution
which results in the minimum shifts and the amplitude deviation of
the form factors of $^{12}$O and $^{28}$S as compared with
$^{16}$O and $^{32}$S. These effects should be observable and show
that elastic electron-nucleus scattering can be used as an
effective tool to study proton drip-line nuclei.

In summary, we have combined the RMF model with the relativistic
partial-wave expansion method to investigate the elastic electron-
or positron-nucleus scattering. Calculations indicate that the
present theory is reliable for the elastic electron- or
positron-nucleus scattering. The elastic electron or positron
scattering cross sections and squared form factors of $^{12, 16}$O
and $^{28, 32}$S are calculated and analyzed. In comparison to
electron, positron scattering from a nucleus displays the minimum
shift and the amplitude deviation of the cross section (in the
same momentum transfer) that is due to the different Coulomb
effect of a nucleus on positron and electron. As compared to the
stable nuclei $^{16}$O and $^{32}$S, both the shifts of the
minimum and the amplitude deviations of the cross section or the
form factors of $^{12}$O and $^{28}$S result from the existence of
the long tail of the charge distribution which is essentially
attributed to the influence of the charge density distribution of
the last two protons in $^{12}$O and $^{28}$S. Since the
difference of the cross sections and form factors between a stable
nucleus and its proton drip-line isotopes is large enough to be
experimentally observable, the elastic electron- or
positron-nucleus scattering is an effective tool to investigate
proton-halo phenomena of proton-rich nuclei.

{}

\end{document}